\documentclass{article}
\usepackage{graphicx}
\usepackage{amsmath}
\usepackage{amsfonts}
\usepackage{amssymb}
\usepackage{bm,bbm}

\begin{document}
\title{{\sl Quartic} quantum theory:\\
       an extension of the standard quantum mechanics}

\author{ Karol \.Zyczkowski$^{1,2}$
\smallskip \\
$^1$Institute of Physics, Jagiellonian University, Krak{\'o}w, Poland\\
$^2$Center for Theoretical Physics, Polish Academy of Sciences, Warsaw,
Poland\\
}

\date{June 30, 2008}

\maketitle

\begin{abstract}
We propose an extended quantum theory, in which the 
number $K$ of parameters necessary to characterize 
a quantum state behaves as {\sl fourth} power
of the number $N$ of distinguishable states.
As the simplex of classical $N$--point probability distributions
can be embedded inside a higher dimensional convex body ${\cal M}_N^Q$ of mixed 
quantum states, one can further increase the dimensionality constructing the
set of extended quantum states. The embedding proposed
corresponds to an assumption that the physical system
described in $N$ dimensional Hilbert space is coupled with an
auxiliary subsystem of the same dimensionality.
The extended theory works for simple quantum systems
and is shown to be a non-trivial generalisation of the 
standard quantum theory for which  $K=N^2$.
Imposing certain restrictions on initial conditions 
and dynamics allowed in the quartic theory
one obtains quadratic theory as a special case. 
By imposing even stronger constraints one arrives at the
classical theory, for which $K=N$.
\end{abstract}

{\small e-mail: karol@tatry.if.uj.edu.pl}

\medskip
\section{Introduction}

For a long time quantum mechanics (QM) 
belongs to the most important cornerstones of modern physics.
Although predictions of quantum theory
were not found to be in contradiction 
with results of physical experiments,
there exist many reasons to look for  
possible generalisations of quantum mechanics--
see e.g. \cite{Gi04,Pe05,Sm06,Sp07, BV07}.

One possible way to tackle the problem 
is to follow an axiomatic approach to quantum mechanics
and to study consequences of relaxing some of the axioms.
An axiomatic approach to quantum theory was 
initiated by Mackey \cite{Ma63},
Ludwig \cite{Lu67,Lu83} and Piron \cite{Pi76}
several decades ago
and further developed in several more recent contributions 
 \cite{Ha01,Mc01,Sc03,Ar07}.
In an influential work of Hardy it is shown  
that quantum mechanics is a kind of probability theory
for which the set of pure states is continuous \cite{Ha01}.
This contrasts with the classical probability theory,
for which pure states form a discrete set of corners of the 
probability simplex. 

Restricting attention to the problem of  finite 
number of $N$ distinguishable states and analyzing
composite systems it is possible
to conclude that the number $K$ of degrees of freedom
(i.e. the number of parameters required to specify a 
given state)
satisfies the relation  $K=N^r$ with an integer exponent.
The linear case, $r=1$, gives the classical 
probability theory. The quadratic case 
leads to the standard quantum theory, for which
it is necessary to use $K=N^2$ real parameters
to characterize completely an unnormalized quantum state.
However, higher order theories may exist, which 
include QM as a special case \cite{Ha01b}.
To single out the standard quantum
theory Hardy uses a 'simplicity axiom' 
and requires that the exponent takes the minimal value
consistent with other axioms, which implies $r=2$. 

Vaguely speaking the exponent $r$ counts the number of indices
decorating a mathematical object called {\sl state},  used to 
determine probabilities associated with the outcomes of a measurement. 
In the classical theory one deals with probability vectors $p_i$
with a single index, while quantum states are described 
by density matrices $\rho_{ij}$. Do we need to work with
some more complicated objects, like tensors or multi-index density matrices
 $\tau_{ijk}$ or $\sigma_{ijkl}$?

The main aim of this work is to propose a higher order theory,
which includes QM as its special case.
Instead of working with higher order tensors, 
the theory of which is well developed \cite{LMV00,Ra02},
we remain within the known formalism of standard complex
quantum mechanics,
and construct an extended quartic (biquadratic)
theory, for which $K=N^4$,
assuming a coupling with an auxiliary subsystem.
The extended quantum mechanics (XM),
reduces to the standard quantum theory
in a special case of uncoupled auxiliary systems.
On the other hand, XM is shown to be a non--trivial generalisation of QM.  
Throughout the entire work only non-relativistic version
of quantum theory will be considered. For simplicity 
we analyze the case of finite dimensional Hilbert space.
Furthermore, we restrict our attention 
to single--particle systems only. 

Well known effects of decoherence reduce the magnitude of quantum effects
and cause a quantum system to behave classically. 
In a similar way one can introduce analogous effects of 'hyper-decoherence'
which cause a system described in the framework of the extended theory 
to lose its subtle properties and behave according to predictions
of standard quantum theory.

It is worth to emphasize that our approach differs 
from the theory  of Adler based on
higher order correlation tensors \cite{Ad07},
the 'two--state vector formalism' of Aharonov and Vaidman \cite{AV91},
the 'time--symmetric quantum mechanics' of Wharton \cite{Wh07},
and  the quaternionic version of quantum theory \cite{Pe79,Ad94,Ad95},
which is not consistent with the power like scaling, $K=N^r$,
in analogy to the quantum theory of real density matrices.

Higher order theory proposed here is also different 
from the algebraic approach of Uhlmann
who discusses spaces of states constructed from Jordan algebra \cite{Uh96},
and from the generalized quantum mechanics developed by Sorkin \cite{So94}.
Furthermore, the extended theory by construction
belongs to the class of {\sl probabilistic} theories 
(see e.g. \cite{BBLW06}), 
so it is different from the theory of
{\sl hidden variables}, which would allow one to predict
an outcome of an individual experiment. 

Our approach explores the geometric structures
in quantum mechanics and in particular the convexity 
of the set of quantum states. Such a description
of quantum mechanics goes back to classical 
papers of Ludwig  \cite{Lu67} and Mielnik \cite{Mi68,Mi69}
and was reviewed and updated in  \cite{BZ06}.

The extended quantum theory constructed here
is related to the generalized quantum mechanics
of Mielnik \cite{Mi74}, in which higher order 
forms on the Hilbert space are considered
and methods of constructing 
non--linear variants of quantum mechanics are discussed. 
On the other hand the theory analyzed here
is linear, and the sets of extended states 
and extended measurements are precisely defined.

The paper is organized as follows.
Section 2 contains a geometric review of the standard set--up of 
quantum mechanics in which we describe the sets of 
quantum states and quantum maps. 
  Discussion of the extended, quartic theory
is based on a definition of the $N^4$ dimensional 
convex set of extended quantum states, introduced 
in sec. 3. In section 4 we describe
the set of generalized measurement operations
admissible in the extended theory while 
section 5 concerns the corresponding
discrete dynamics.
Section 6 contains the evidence
that the extended theory forms
a non-trivial generalisation of the standard, quadratic quantum theory.
The possibility of generalizing the quantum theory even further
and working with higher order theories is discussed 
in section 7. The work is concluded with a discussion 
in section 8, while some information on duality between 
convex sets is presented in Appendix A.


\section{Standard quantum theory:  quadratic}

In this section we review the standard quantum theory
and present requisites necessary for its generalisation.
Quantum mechanics is a probabilistic theory.
Probabilities associated with outcomes of a measurement
are characterized by a quantum state
described by a density operator $\rho$
which acts on $N$--dimensional Hilbert space ${\cal H}_N$.
In this work we shall assume that $N$ is finite.
The density operator is Hermitian and positive.

The set of normalized quantum states of size $N$
for which Tr$\rho=1$ will be denoted by ${\cal M}_N^Q$.
In the simplest case of $N=2$ the
set of mixed states of a single qubit forms the {\sl Bloch ball},
${\cal M}_2^{Q}=B_3\subset {\mathbb R}^3$.
Degree of mixing of a state $\rho$ can be characterized by the
von Neumann entropy, $S(\rho)=-{\rm Tr }\rho \ln \rho$.
This quantity varies from zero for pure states, to $\ln N$
for the maximally mixed state, $\rho_*={\mathbbm 1}/N$,
located in the center of the set  ${\cal M}_N^Q$.

To introduce a partial order into the set of mixed states 
one uses the majorization relation \cite{MO79}.
A density matrix $\rho$ of size $N$
is majorized  by a state $\omega$, 
written $\rho \prec \omega$,
if their decreasingly ordered spectra $\vec \lambda$ and $\vec \kappa$ satisfy:
$\sum_{i=1}^m \lambda_i \le \sum_{i=1}^m \kappa_i$,
for $m=1,2,\dots, N-1$.
The majorization relation implies an inequality between entropies:
if  $\rho \prec \omega$ then $S(\rho) \ge S(\omega)$.
Any mixed state $\rho$ satisfies relations 
$\rho_*  \prec \rho \prec|\psi\rangle\langle \psi|$,
where $|\psi\rangle\in {\cal H}_N$ 
denotes an arbitrary pure state -- see e.g. \cite{BZ06}.

For our purposes it is also
convenient to work with  {\sl subnormalized states}, such that 
Tr$\rho \le 1$. The  $N^2$ dimensional set 
of subnormalized states forms a convex hull
of the set of normalized states and the zero state, 
 ${\widetilde{\cal M}}_N^{Q}={\rm conv  ~  hull }
   \{  {\cal M}_N^Q ,  0 \}$.

A one--step linear dynamics in  ${\cal M}_N^Q$ 
may be represented in its Kraus form
\begin{equation} \rho \: \rightarrow \: 
\rho' \ = \ \Phi(\rho)  \: = \: \sum_{i=1}^k X_i\,\rho\, 
X^{\dagger}_i \ ,
\label{Kraus1}
\end{equation} 
in which the number $k$ of Kraus operators can be arbitrary.
Such a form ensures that the map $\Phi$ is 
{\sl completely positive} (CP), which means that an extended
map, $\Phi \otimes {\mathbbm 1}_M$, 
sends the set of positive operators into itself
for all possible dimensions $M$ of the ancilla \cite{Kr71}.
The Kraus operators $X_i$ can be interpreted
as measurement operators,  
and the above form provides a way
to describe quantum measurement performed on the state $\rho$:
The $i$--th outcome occurs with the probability
$p_i={\rm Tr}X_i \rho X_i^{\dagger}$
and the measurement process transforms the initial state according to 
\begin{equation}
 \rho \: \rightarrow   \rho_i  \ = \frac{X_i\,\rho\, X^{\dagger}_i}{{\rm Tr}X_i
  \rho X_i^{\dagger}} \ .
\label{Krausib}
\end{equation} 

To assure that the trace of $\rho$ 
does not grow under the action of $\Phi$,
 the Kraus operators $X_i$ 
need to satisfy the following inequality \cite{Kr71}, 
\begin{equation} \sum_{i=1}^k X^{\dagger}_i X_i \: \le 
\: {\mathbbm 1}_N \ .
\label{Kraus2} 
\end{equation}  

Usage of subnormalized states and trace non--increasing maps
corresponds to a realistic physical assumption
that the experimental apparatus fails to work
with a certain probability and no measurement results are recorded.

The measurement process can be characterized by 
the elements of POVM (positive operator valued measures),
$E_i:=X^{\dagger}_i X_i$. By construction these operators
are hermitian and positive. Due to (\ref {Kraus2}) they fulfill 
the relation $\sum_i  E_i \le {\mathbbm 1}_N$
hence each individual element satisfies $E_i \le {\mathbbm 1}_N$.
Thus the set of the elements
of a POVM in the standard quantum theory can be defined as 
%
%
\begin{equation} 
{\cal E}_N^Q \ := \ \{
E_i=E_i^{\dagger}: E_i\ge 0 {\rm \quad and \quad} E_i
\le {\mathbbm 1}_N  \}  \ .
\label{povm1} 
\end{equation}
Since the elements of POVM are positive operators, 
the probability $p_i$ is non--negative,
\begin{equation} 
 p_i= {\rm Tr}\, \rho E_i \ge 0
 {\rm \quad for \quad any \quad}
 \rho \ge 0  \ . 
\label{trace1} 
\end{equation}
The above relation shows that the cone containing the elements of POVM 
is {\sl dual} to the set of subnormalized states, 
${\cal E}_N^Q=({\widetilde{\cal M}}_N^{Q})^{*}$.
In the case of the standard quantum theory
we work with the set of positive operators which is selfdual, 
so both cones are equal, 
${\cal E}_N^Q={\widetilde{\cal M}}_N^{Q}$ -- see Fig \ref{fig4}a.
For more information on dual cones consult appendix A.

In the special case of equality in (\ref{Kraus2})
the completeness relation  
$\sum_i X^{\dagger}_i X_i = {\mathbbm 1}_N$
imposes that Tr$\Phi(\rho)= {\rm Tr} \rho$.
A completely positive trace preserving map
is called {\sl quantum operation} or {\sl stochastic} map.
If the dual relation is satisfied, 
$\sum_i X_i  X^{\dagger}_i= {\mathbbm 1}_N$,
 the map is called {\sl unital},
since it preserves the maximally mixed state,
$\Phi(\rho_*)=\rho_*={\mathbbm 1}/N$.
A completely positive trace preserving unital 
map is called {\sl bistochastic}.
We are going to use an important property of these maps
reviewed in \cite{BZ06}:
Any initial state majorizes its image
with respect to any bistochastic map,
$\Phi_B(\rho) \prec \rho$.

Treating $\rho$ as an element of the Hilbert-Schmidt space of operators,
we may think of $\Phi$ as a 
super--operator,
(a square matrix of size  $N^2$), acting in this space,
\begin{equation}  
\rho' _{m\mu}  \: = \: 
 \Phi_{\stackrel{\scriptstyle m \mu}{n \nu}}
  \, \rho_{n \nu} \ ,
\label{map2}
\end{equation} 
where summation over repeated indices has to be performed.
The operators acting on  the vectors of Hilbert-Schmidt
space are often called {\sl super--operators}, in order
to distinguish them from the operators of HS space itself.
Let us emphasize that this common notion \cite{Mu02} 
used since the sixties \cite{Pr63} is not related
to supersymmetric theories.
 
The superoperator $\Phi$ can be represented 
 by means of tensor products of the Kraus operators,
\begin{equation}  
\Phi= \sum_{i=1}^k X_i \otimes {\bar X_i} \ .
\label{supermap}
\end{equation} 
The matrix $\Phi$ needs not to be Hermitian.
However, reshuffling its elements one defines 
a  Hermitian  {\it dynamical matrix}, $D=D(\Phi)$ \cite{SMR61}.
Its elements read
\begin{equation}
D_{\stackrel{\scriptstyle m n}{\mu \nu}} : =
\Phi_{\stackrel{\scriptstyle m \mu }{n \nu}} =
(\Phi_{\stackrel{\scriptstyle m  n }{\mu  \nu}})^R  \ . 
 \label{dynmatr3}
\end{equation}
The symbol $^R$ denotes the   
transformation of reshuffling of elements of a four-index matrix, 
which exchanges two indices, $\mu$ and $n$ in the formula above
\cite{BZ06}.

A  theorem of Choi \cite{Cho75a} states that the map $\Phi$
is completely positive, if the dynamical matrix  is positive,
$D(\Phi)\ge 0$.
Therefore  $D$, also called {\sl Choi} matrix, might be interpreted
as a Hermitian operator acting on a composed Hilbert space
${\cal H}_{N^2}={\cal H}_A \otimes {\cal H}_B$
of size $N^2$. The trace non-increasing condition
(\ref{Kraus2}) is equivalent to the 
following constraint on the dynamical matrix,  
Tr$_A D \le  {\mathbbm 1}_N$.
 Rescaling the Choi matrices $\sigma=D/N$,
we recognize that the set of trace non--increasing maps
forms a convex subset of the $N^4$ dimensional
set of subnormalized states on ${\cal H}_{N^2}$ \cite{CSZ07}.
This  duality between linear maps 
and states on the enlarged system is called 
{\sl Jamio{\l}kowski isomorphism}, 
which refers to his early contribution \cite{Ja72}. 

To appreciate this duality let us look at an extended operation
$(\Phi \otimes {\mathbbm 1})$ acting on the maximally entangled state
\begin{equation}  
|\psi^+\rangle \ = \ \frac {1}{\sqrt{N}} 
 \sum_{i=1}^N |i\rangle \otimes |i\rangle 
\label{entangmax}
\end{equation} 
from the enlarged Hilbert space, ${\cal H}_N \otimes {\cal H}_N$.
The dynamical matrix $D$
 corresponding to the map $\Phi$ reads then 
\begin{equation}  
 D(\Phi) = N(\Phi \otimes {\mathbbm 1})|\psi^+\rangle \langle \psi^+| \ .
\label{jamiol2}
\end{equation} 
The state--map isomorphism, written above for states 
with maximally mixed partial trace, Tr$_A D =  {\mathbbm 1}_N$, 
can be generalized also for other states  \cite{Le06}.

After reviewing some basic properties of 
discrete quantum dynamics, let us see how 
classical dynamics emerges as a special case
of the quantum theory.
Consider the set of normalized 
diagonal density matrices, $\rho_{ij}= p_i \delta_{ij}$,
which forms the $(N-1)$-dimensional simplex ${\cal M}_N^C=\Delta_{N-1}$
of classical probability distributions.
If we restrict our attention to
maps described by diagonal dynamical matrices, 
$D_{\stackrel{\scriptstyle m \mu}{n \nu}} = T_{m\mu}
\delta_{mn}\delta_{\mu \nu}$, 
then the diagonal structure of $\rho$ is preserved,
so we recover the classical probability theory.
Moreover, for any quantum map $\Phi$
we obtain corresponding classical dynamics in
the simplex of probability distributions,  
$p'_m=T_{mn}p_n$, by constructing the
transition matrix out of diagonal elements of 
the dynamical matrix.

\medskip
{\bf Lemma 1}. {\sl Let $\Phi$ be a linear quantum map acting on 
${\cal M}_N^Q$. 
Let $T$ denote a square matrix of size $N$
obtained by reshaping the diagonal elements of the corresponding
dynamical matrix, 
$T_{mn}= \Phi_{\stackrel{\scriptstyle mm}{nn}}$, 
(without summation over repeating indices).
If $\Phi$ is a quantum stochastic (bistochastic) map,
then $T$ is a stochastic (bistochastic) matrix.}
\medskip

{\bf Proof}. If quantum map $\Phi$ is completely positive,
the corresponding dynamical matrix is positive definite,
so all elements of its diagonal used to assemble $T$
are not negative. If $\Phi$ is trace preserving,
equality in (\ref{Kraus2}) holds, 
and implies the relation $\sum_n T_{mn}=1$
for all $m=1,\dots, N$. 
If $\Phi$ is unital then the dual relation for 
partial trace of $\Phi^R$ implies that 
 $\sum_m T_{mn}=1$ for all $n=1,\dots, N$. 
Thus quantum stochasticity (bistochasticity)
of the map $\Phi$ implies the classical 
property of the transition matrix $T$. $\square$
\medskip

Any trace non increasing map,
 for which the Kraus operators satisfy relation (\ref{Kraus2}),
 can be also called {\sl sub--stochastic},
 since in the case of diagonal $D$
 the classical transition matrix $T$ is sub--stochastic \cite{MO79}.

\begin{figure} [htbp]
      \begin{center} \
  \includegraphics[width=12.0cm,angle=0]{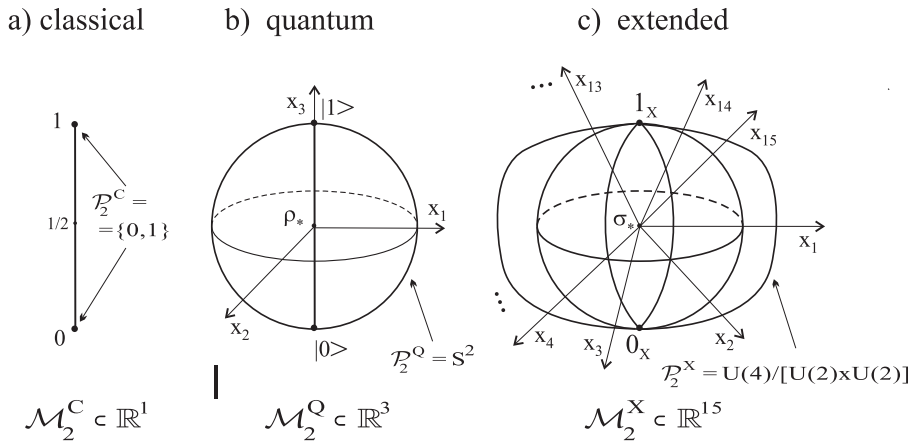}
\caption{The space of mixed states for $N=2$: a) 
classical theory,  b) quantum theory, c) extended theory (sketch of a $15$--d
set).}
\label{fig1}
\end{center}
    \end{figure}

It is instructive to analyze the geometry of the sets
of classical and quantum states \cite{BZ06}.
For simplicity we have compared in Fig. 1
the sets of normalized states for $N=2$.
The interval of classical one--bit states ${\cal M}_2^C=[0,1]$
can be embedded inside the 3-d Bloch ball
consisting of pure and mixed states of one qubit  -- see Fig. 1b.
On the other hand, the Bloch ball
can be inscribed inside the cube, which describes
the set of uncorrelated states of
three classical bits -- see Fig. 2a. This 
very cube can be  embedded inside the simplex $\Delta_7$
formed by $2^3$ corners representing all possible
pure states of three classical bits. 
The cube  would describe allowed results of 
fiducial measurements of three components of the spin $1/2$,
if they were independent classical variables \cite{Ba06}.
Truncation of the corners of the $3$--cube,
implied by rules of quantum mechanics,
reduces the cube to the ball.
After this truncation procedure only two
quantum states remain distinguishable.
Furthermore, such a qualitative change of the
 symmetry of the body 
makes the set of extremal states continuous and 
allows for an arbitrary rotation of the Bloch ball \cite{Ha01b, Ga05}.
Rotating the initial state $|0\rangle$
with respect to any axis perpendicular to 
the interval of classical states one generates
a coherent superposition of $|0\rangle$
and $|1\rangle$. Existence of such a pure state,
which does not have a classical analogue,
explains interference effects, typical of quantum theory.

\section{An extended quantum theory:  quartic}

In order to work out a generalized, quartic theory
we need first to define a set of extended states.
In this paper we are going to consider mono--partite systems\footnote
{Mono-partite systems consist of a single particle only, 
while bi-partite systems consist of tow well defined subsystems.}
of an arbitrary size $N$, but to gain some intuition
we shall begin with the simplest case of $N=2$.
Analyzing the normalized states of a single 
{\sl qubit} (quantum bit),
we will copy the embedding procedure which 
blows up the interval ${\cal M}_2^C$ of classical states
into the Bloch ball ${\cal M}_2^Q$ of quantum states. 
Thus we shall put the $3$--d Bloch ball of all states of a qubit
inside the larger $15$--d body ${\cal M}_2^X$ of an {\sl exbit}
(extended bit) as sketched in Fig 1c.
(In a recent paper of Barrett \cite{Ba06}
a similar name  of {\sl gbit}, standing for {\sl generalized bit}
was introduced). 
Designing the shape of the set of extended bits 
 we have to keep in mind
that it may contain only two distinguishable states,
say $0_X$ and $1_X$. 

The dimension of the set ${\cal M}_N^X$ of normalized extended states
should be equal to $N^4-1$, since the remaining dimension
is obtained by imposing a weaker condition of  subnormalization.
Thus  it is natural to look at it as a
suitable subset of the set  ${{\cal M}}_{N^2}^{Q}$
which contains the mixed states of two {\sl quNits} 
(systems described in $N$-dimensional Hilbert space).
As the ball of one--qubit states 
arises by truncating the corners of the three--bit cube,
to define the set of the states of an exbit
we propose to reduce the number of distinguishable states 
in  ${{\cal M}}_{4}^{Q}$ by 
truncating the corners of the simplex $\Delta_3$
of eigenvalues of standard quantum states for $N=4$.
In this way one obtains a convex $15$-d set of these states, 
the spectra of which belong to the octahedron
formed by $6$ centers of edges of the tetrahedron $\Delta_3$ - see Fig 2b.

More formally,  let us propose a general definition of the set
of extended states for an arbitrary $N$,
 \begin{equation} 
{{\cal M}}_{N}^{X} := \{\sigma\in {\cal M}_{N^2}^Q:  
 \sigma \prec \sigma_{0}:=
|0\rangle \langle 0 | \otimes \frac{1}{N}{\mathbbm 1} \} \ ,
\label{restrict1} 
\end{equation} 
where $|0\rangle \in {\cal H}_N$
represents an arbitrary pure state of the standard theory.
The symbol $\prec $ denotes the majorization relation
(defined in previous section),
with the help of which the truncation is performed.
The ordered set of eigenvalues of the extended state $\sigma_0$ 
forms the vector of length $N^2$ with $N$ non-vanishing components only,
 eig$(\sigma_{0})=\frac{1}{N}\{ 1, \dots,1,0,\dots,0\}=:{\vec v}$.
From a combinatorial point of view
the set of all spectra majorized by this vector 
forms a permutation polytope  called {\sl  permutohedron}
or multipermutohedron \cite{On93,BS96}.
It is defined as a convex hull
of all permutations of a given vector,
${\rm Perm}({\vec v}):={\rm conv hull}\bigl( P({\vec v}) \bigr)$,
where the convex hull contains all $k!$ permutations $P$ in the $k$--element set
of components of  ${\vec v}$.
In the case considered here
the vector $\vec v$ has $k=N^2$
components, but only $N$ of them are non--zero.
Thus the number of corners  of ${\rm Perm}_N:={\rm Perm}({\vec v})$
is given by the binomial symbol, $C_N= \binom{N^2}{N}$.
In the simplest case of $N=2$ we get
$C_2=6$ and the set ${\rm Perm}_2$ forms a regular octahedron  shown in Fig. 2b.
Thus an operator $\sigma$ belongs to the
set of extended states
 if its spectrum belongs to the permutohedron,
\begin{equation} 
{{\cal M}}_{N}^{X} = \{\sigma=\sigma^{\dagger}: 
{\rm eig}(\sigma) \in {\rm Perm}_N \} \ .
\label{restrict2} 
\end{equation} 

\begin{figure} [htbp]
      \begin{center} \
  \includegraphics[width=12.0cm,angle=0]{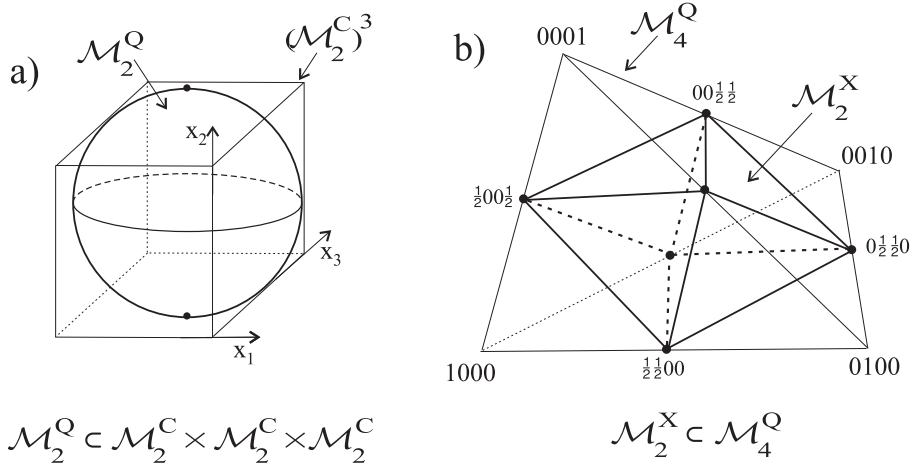}
\caption{Embedding of a) the Bloch ball of one qubit states inside the cube 
 of states of three independent bits, b) the octahedron
  of one exbit states inside the tetrahedron of two--qubit states 
  (to reduce the dimensionality only spectra
are considered).}
\label{fig2}
\end{center}
    \end{figure}

To show that the described choice of the set ${{\cal M}}_{N}^{X}$ is acceptable
we need to discuss the number of distinguishable states it supports.

\medskip
{\bf Lemma 2}. {\sl The set ${{\cal M}}_{N}^{X}$ contains 
exactly $N$ mutually distinguishable states.}
\medskip

{\bf Proof}. 
 Let $\{|i\rangle\}_{i=1}^N$
represent an arbitrary orthonormal basis in ${\cal H}_N$.
Then extended states $\sigma_i=|i\rangle \langle i | \otimes{\mathbbm 1}/N $ 
have non---overlapping supports and  can be deterministically discriminated.
To show that ${{\cal M}}_{N}^{X}$ does not contain
more distinguishable states it is sufficient to apply a lemma, that the sum of the
ranks of all distinguishable states is not larger than
the total dimension $d_t$ of the Hilbert space \cite{MMZ07}.
 In the case analyzed  $d_t=N^2$ and the set ${{\cal M}}_{N}^{X}$
does not contain any states with rank smaller than $N$,
so the maximal number of distinguishable states is  equal to $N$  $\square$.
\smallskip

By means of a suitable mixture of unitary transformations one
can send the state $\sigma_{1}$  into any state $\sigma$
such that $\sigma\prec \sigma_{1}$. A convex mixture of unitaries 
is bistochastic,
so any initial state majorizes its image obtained with this map. 
Thus both conditions are equivalent and we are in a position
to formulate an alternative definition of the set of extended states,
\begin{equation} 
{{\cal M}}_{N}^{X} = {\rm conv \ hull} \bigl[ U \bigl(
 |0\rangle \langle 0 |  \otimes \frac{1}{N}{\mathbbm 1}
  \bigr) U^{\dagger} \bigr] ,
\label{restrict1b} 
\end{equation} 
where $|0\rangle$ is an arbitrary state in ${\cal H}_N$, 
while $U$ denotes a unitary matrix from $U(N^2)$.

Let us note that the set ${{\cal M}}_{N}^{X}$
of extended states defined in equivalent forms
({\ref{restrict1}, \ref{restrict2}) and (\ref{restrict1b})
is determined as the minimal set in $N^4-1$ dimensions 
which  is invariant under unitary transformations
and supports exactly $N$ distinguishable states.

In the simplest case of $N=2$ the set ${{\cal M}}_{2}^{X}$ 
of extended states has an appealing property. By 
construction it forms a convex subset of the set of two--qubit
states, which contains two sets of separable and entangled states.
Consider a three dimensional subset of the set of two--qubit states
defined as convex combination of four Bell states,
$|\psi_{\pm}\rangle =(|00\rangle \pm |11\rangle)/\sqrt{2}$ and 
$|\phi_{\pm}\rangle =(|01\rangle \pm |10\rangle)/\sqrt{2}$.
Then the octahedron contained in ${{\cal M}}_{2}^{X}$ 
consists of separable states only, while all other states
of the tetrahedron are entangled.

For $N=2$ any pure state of the extended  theory 
is represented by an $N=4$ mixed state of the standard theory 
$\sigma_{\phi}= |\phi\rangle \langle \phi |  \otimes \rho_*$
with spectrum $\{1/2,1/2,0,0\}$. The orbit of 
pure states of the extended theory
contains  unitarily equivalent states,  
which form an $8$-dimensional flag manifold,
${\cal P}_2^X=U(4)/[(U(2)\times U(2)]$
in contrast to the $2$ dimensional Bloch sphere, 
${\cal P}_2^Q=U(2)/[(U(1)\times U(1)]={\mathbbm C}P^1=S^2$.
In general,
 the set of pure states of the extended theory,
${\cal P}_N^X=U(N^2)/[(U(N^2-N)\times U(N)]$,
has $2N^2(N-1)$ dimensions. This is exactly $N^2$ times 
more than its quantum counterpart,  
the complex projective space of $2(N-1)$ real dimensions, 
${\cal P}_N^Q=U(N)/[(U(N-1)\times U(1)]={\mathbbm C}P^{N-1}$.
By construction the entropy of an extended state belongs to the interval
$S(\sigma)\in [\ln N, 2\ln N]$,
so it is convenient to define a gauged quantity
$S_X:=S-\ln N$ which vanishes for extended pure states.

To find out how the Bloch ball is embedded inside ${\cal M}_{2}^{X}$ consider 
a family of one qubit states $\rho(a)$ with spectrum $\{a,1-a\}$. 
The extension of this family has the form
$\sigma(\rho)=\rho \otimes {\mathbbm 1}/2$.
The spectra of these extended states read  $\{1/2-a,1/2-a,a,a \}$
and form the vertical diagonal of the octahedron
which crosses its center and joins the
edges $\{1/2,1/2,0,0\}$ with  $\{0,0,1/2,1/2\}$ (see Fig. 2b).
These points represent the logical states of the extended theory, 
$0_X$ and $1_X$, equal to $|0\rangle \langle 0| \otimes \rho_*$  and 
$|1\rangle \langle 1| \otimes \rho_*$ respectively.
All other points of the Bloch ball are obtained
from points of the interval 
$[0_X, 1_X]$
by local unitary rotation, $V\otimes {\mathbbm 1}$.
Note that the points of the octahedron ${\cal M}_{2}^{X}$ 
beside the vertical interval $[0_X ,1_X]$
do not have one--to--one analogues in the quantum theory.

In the case of an arbitrary $N$ 
an extension of $\rho$ is obtained by 
adding an ancilla in the maximally mixed state, 
\begin{equation} \rho \to  \sigma \equiv
\rho \otimes \frac{1}{N}\,{\mathbbm 1}_{N}
{\rm \quad \quad hence \quad \quad}
\sigma_{ijkl}=\frac{1}{N}\,\rho_{ik}\delta_{jl}
\label{extend} 
\end{equation}   
By construction these states belong to ${\cal M}_{N}^{X}$
and act on an extended Hilbert space
${\cal H}_{N^2}={\cal H}_{A}\otimes {\cal H}_{A'}$.
Moreover, pure states of the standard quantum theory with vanishing entropy  
are mapped into extremal states of the extended theory
with entropy equal to $\ln N$.

In general a state $\sigma$ of the extended theory
need not have the product form (\ref{extend}),
since the state $\rho$ may be 
entangled with the ancilliary system. 
A bipartite state $\sigma\in {\cal M}_N^X$ 
will be called an {\sl extension} 
of the quantum state $\rho\in {\cal M}_N^Q$ 
if the marginal satisfies 
\begin{equation} 
{\rm Tr}_{A'} (\sigma)  \ = \ \rho  \ .
\label{traceb}
\end{equation}  
Reduction by partial trace is not reversible,
and a given mixed state $\rho$ may have several different
extensions $\sigma$, such that ${\rm Tr}_{A'} (\sigma) =\rho$.
However, any pure state has a unique extension only. 
It has a tensor product form (\ref{extend}) and reads 
$\sigma_{\phi}=
|\phi\rangle \langle \phi | \otimes \frac{1}{N}{\mathbbm 1}$.

Extended states $\sigma \in {\cal M}_{N}^{X}$
are defined on an bipartite system
and can be interpreted in view of 
the Jamio{\l}kowski isomorphism (\ref{jamiol2}):
A state $\sigma$ of the extended theory 
may be considered as a completely positive 
quantum map acting on ${\cal M}_{N}^{Q}$ 
and determined by $D=N\sigma$.
In general, a map need not be trace preserving,
since this is only true if Tr$_A\sigma={\mathbbm 1}/N$.

In particular this is the case for all product extensions
(\ref{extend}) for which 
$D=\rho\otimes {\mathbbm 1}$.
The corresponding map $\Phi_{\rho}$
acts as a complete one--step contraction,
and sends any initial state $\omega$ into $\rho$,
\begin{equation}
\Phi_{\rho}(\omega)=\rho 
{\rm \quad for \quad any \quad}
\omega \in {\cal M}_N^Q \ .
\label{extend2} 
\end{equation}  
To show this let us start with the dynamical matrix of this map,  
$D_{\stackrel{ \scriptstyle m n}{\mu \nu}}=\rho_{m\mu}\delta_{n \nu}$. 
Writing down the elements of the image 
$\omega'=\Phi_{\rho}(\omega)= D^R \omega$
in the standard basis we obtain the desired result,
$\omega'_{m\mu}=D_{\stackrel{ \scriptstyle m n}{\mu\nu}}\; \omega_{n \nu}= 
\rho_{m\mu} ({\rm Tr}\; \omega )=  \rho_{m\mu}$.

Consider now an arbitrary state $\sigma \in {\cal M}_{N}^{X}$, 
prepared as an extension of $\rho={\rm Tr}_B \sigma$.
This relation can be rewritten with help 
of the superoperator (\ref{supermap}), dynamical matrix (\ref{dynmatr3})
and Jamio{\l}kowski isomorphism (\ref{jamiol2}),
\begin{equation} 
\rho={\rm Tr}_B \sigma = \frac{1}{N}  \sum_{i=1}^k X_i X_i^{\dagger}=
\Phi({\mathbbm 1}/N) \ . 
\label{contr2} 
\end{equation}  
In this way we have arrived at a dynamical 
interpretation of objects of the extended theory:
A state $\rho \in {\cal M}_{N}^{Q}$ 
can be extended to $\sigma \in {\cal M}_{N}^{X}$, 
which represents a linear map $\Phi:{\cal M}_{N}^{Q} \to {\cal M}_{N}^{Q}$, 
such that its effect is equal to $\rho$.
It means that $\Phi$
sends the maximally mixed state $\rho_*={\mathbbm 1}/N$
into $\rho$.
 In particular the trivial (product) extension 
(\ref{extend}) represents the complete contraction, which sends 
every initial state into $\rho$. 

In analogy to the $N^2$ dimensional set
${\widetilde{\cal M}}_{N}^{Q}$
of quantum subnormalized states
we define subnormalized extended states, 
satisfying Tr$\, \sigma\le 1$.
 The set of all such states,  
 ${\widetilde{\cal M}}_{N}^{X}={\rm conv  ~  hull }
  \{  {\cal M}_N^X ,  0 \}$
has $N^4$  dimensions, as required.
Thus the number $K$ of parameters necessary to characterize 
a given state $\sigma$  
of the theory behaves as the forth power of $N$.
This property justifies the name
used in the title of the paper:
the extended theory proposed
in this work can be called {\sl quartic}.

Let us compare our construction with the
generalized quantum mechanics of Mielnik \cite{Mi74}.
In his approach the set of extended states contains
'density tensors' constructed of convex combinations
of  separable pure states.
On the other hand the set of extended states ${\cal M}_{N}^{X}$
contains also states which are not separable 
with respect to the fictitious splitting into 
the 'physical system' and the 'hypothetical ancilliary system'.
Such a possible entanglement between the 
'system' and the 'ancilla' plays a crucial role in the 
dynamics: as shown in the subsequent sections
it contributes to the fact that the predictions 
of the standard quadratic theory
and the generalized quartic theory can be different.

To summarize this section, a quartic extension 
of the quadratic quantum theory
is constructed by extending the set of admissible states.
Any extended state $\sigma$ can be interpreted as if the  
corresponding state $\rho$ of the standard theory were 
entangled with an auxiliary  subsystem of the same size in such a way that
the state $\sigma$ of the composite system  
obeys (\ref{restrict1}) and its marginal is equal to $\rho$.
Note that such a concept of an ancillary subsystem
(the presence of which might be difficult to detect)
is introduced for a pedagogical purpose only:
In the extended theory the system is 
described by a single density tensor with four indices,
so in practice it cannot be divided into 
a 'physical particle' and an  auxiliary 'ghost-like' subsystem.

A state $\sigma$ of the extended theory 
represents a completely positive quantum map which
moves the center of the body of quantum states into $\rho$.
As the set of classical states forms the
set of diagonal density matrices embedded in
${\cal M}_{N}^{Q}$, the set of quantum states
forms a proper subset of ${\cal M}_{N}^{X}$
containing product states,
$\rho\otimes {\mathbbm 1}/N$. 

\section{Extended measurements and POVMs}
\label{sec:POVM}

In analogy to the standard quantum theory of a
measurement process we will assume that the generalized measurement
acting on $\sigma$
is described by the elements $E^X_i$
of an extended POVM. 
Conservation of probability implies
a relation $\sum E^X_i \le 1$
in analogy to the quantum case.

Furthermore, the probabilities $p_i$ of a single outcome
have to be non--negative, so we require that
\begin{equation} 
 p_i= {\rm Tr}\, \sigma E^X_i \ge 0
 {\rm \quad for \quad any \quad}
 \sigma \in {\cal M}_N^{X} \  .
\label{trace2} 
\end{equation}
This equation defines the set of 
elements $E_i^X$ of an XPOVM (an extended POVM). 
The key difference
with respect to the quantum condition (\ref{trace1})
is that the extended states $\sigma$ are not only positive,
but they belong to the set ${\cal M}_N^{X}$
which arise by truncation of the set ${\cal M}_{N^2}^{Q}$
of positive operators acting on an extended system.
Hence  elements $E_i^X$ of an XPOVM (an extended POVM) 
may not be positive, provided the condition (\ref{trace2}) holds for all
admissible states. This relation shows that the
set of elements of XPOVM  belongs to the cone {\sl dual} 
to the set of extended states, 
\begin{equation} 
{\cal E}_N^X \ := \ \{
E^X_i=(E^X_i)^{\dagger}: E^X_i \in ({\cal M}_N^{X})^{*}
 {\rm \quad and \quad} E^X_i \le {\mathbbm 1}_N  \}  \ .
\label{povme1} 
\end{equation}
Here ${\cal M}^*$ denotes the set dual to ${\cal M}$ 
-- see Appendix A for definition and properties of 
dual cones and dual sets.

\begin{figure} [htbp]
      \begin{center} \
  \includegraphics[width=11.7cm,angle=0]{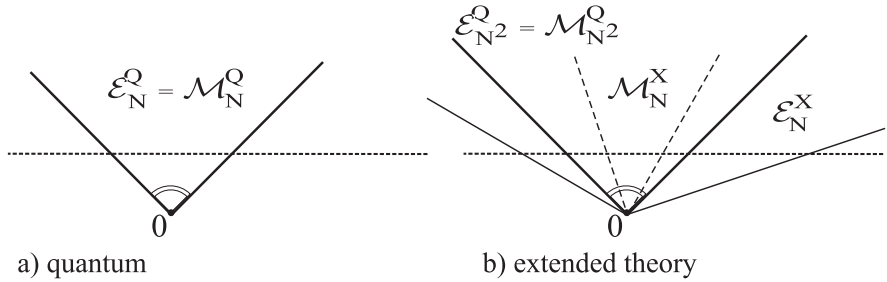}
\caption{Cones of states ${\cal M}$ and elements of POVM  $\cal E$:
a) selfdual for quantum theory,
${\cal E}_{N}^Q= {\cal M}_{N}^Q$;
b) dual in the extended theory for which   
 ${\cal E}_N^X \supset {\cal E}_{N^2}^Q= {\cal M}_{N^2}^Q
 \supset {\cal M}_N^X=({\cal E}_N^X)^*$.}
\label{fig3}
\end{center}
    \end{figure}

The geometry of these sets is sketched in Fig \ref{fig3}.
While in the case of standard quantum theory both sets of quantum states
states and elements of POVM do coincide (panel a),
in the extended theory the set ${\cal M}_N^{X}$
of extended states does not contain
all positive operators, so the dual set  
${\cal E}_N^X$ contains also some 
operators which are not positive.
The boundary of the cone of  the elements of POVM 
has to be perpendicular to the opposite boundary 
of the set of extended states, 
since the relation (\ref{trace2}) 
bounds the scalar product in the Hilbert--Schmidt space
of linear operators. Thus this relation 
can be interpreted as a condition that the angle between two 
corresponding vectors is not larger than $\pi/2$.

\begin{figure} [htbp]
      \begin{center} \
  \includegraphics[width=10.0cm,angle=0]{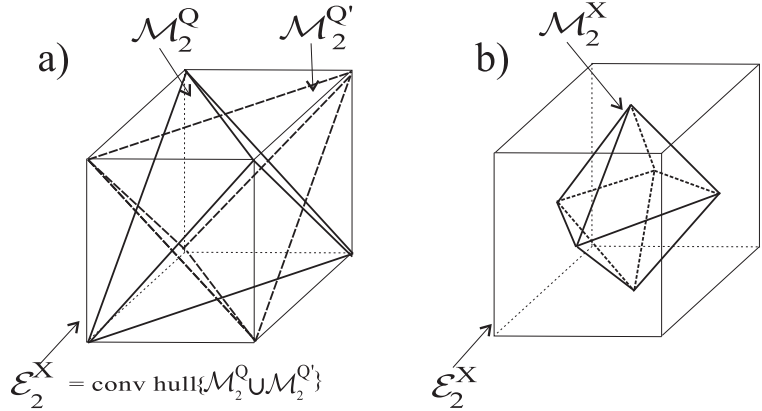}
\caption{The space of elements of an extended POVM
forms a)  a convex hull of the set  ${\cal M}_2^Q$ of states and its
image and
b) set  ${\cal E}_2^X$ dual to the set  ${\cal M}_2^X$ of extended states.}
\label{fig4}
\end{center}
    \end{figure}

Since the set of extended states is invariant with respect
to unitary transformation, its structure is determined 
by the permutohedron ${\rm Perm}_N$ containing
all admissible spectra. Therefore the set  
${\cal E}_N^X$ of all elements of XPOVM
is unitarily invariant and can be specified
by defining its spectra. Lemma  4 proved in appendix A
implies that   
\begin{equation} 
{\cal E}_N^X \ := \ \{
E^X_i=(E^X_i)^{\dagger}: \ 
 {\rm eig} ( E^X_i)  \in ( {\rm Perm}_N)^{*}
 {\rm \quad and \quad} E^X_i \le {\mathbbm 1}_{N^2}  \}  \ .
\label{povme2} 
\end{equation}

Hence to find the set of operators belonging to ${\cal E}_N^X$
it is  sufficient to find a polytope dual to the permutohedron
${\rm Perm}_N$.
Each corner of the permutohedron ${\rm Perm}_N$
generates a face  of $({\rm Perm}_N)^{*}$, so 
the latter polytope has  $C_N= \binom{N^2}{N}$ faces.
The structure of both sets is particularly simple in case of $N=2$, for which 
one arrives at a pair of dual regular polytopes in ${\mathbbm R}^3$:
an octahedron and a cube. The spectra of the states from 
${\cal M}_2^{X}$ belong to the regular octahedron 
${\rm Perm}_2={\rm Perm}(1/2,1/2,0,0)$,
so the dual set ${\cal E}_2^X$ of elements
of XPOVM contains operators with spectra belonging to the
cube equal to $({\rm Perm}_2)^{*}$. The cube
can be written as convex hull
of a tetrahedron and its mirror copy,
$({\rm Perm}_2)^{*}= {\rm conv \ hull}
\Bigl( {\rm Perm}(1,0,0,0)\;  \cup  \; {\rm Perm}(1/2,1/2,1/2,-1)\Bigr)$ --
see Fig.  \ref{fig4}.
Observe that ${\cal M}_2^{X}$ contains also non--positive
matrices. e.g. a diagonal matrix ${\rm diag}(1/2,1/2,1/2,-1)$.
However, due to duality  relation
${\cal E}_N^X  =\bigl({\cal M}_N^X\bigr)^{*}$,
inequality (\ref{trace2}) is by construction fulfilled
for any extended state $\sigma \in  {\cal M}_N^X$.

\section{Extended dynamics and super-maps}

To complete the construction of the quartic theory 
we have to allow for some action in the set
of extended states. As in the case
of standard quantum theory we shall discuss 
only discrete linear maps.

An extended state $\sigma \in {\cal M}_{N}^{X}$ 
may be considered as a map on  ${\cal M}_{N}^{Q}$,
so we are going to analyze a transformation
$\sigma'=\Gamma(\sigma)$
which sends a quantum map into a quantum map.
Since in the physics literature a map sending operators into operators is
called a 'super--operator',
we shall call $\Gamma: {\cal M}_{N}^{X} \to {\cal M}_{N}^{X}$
a {\sl super-map}. 
It is represented by a matrix of size $N^4$
which  acts on an extended state in analogy to (\ref{map2}),
\begin{equation}  
\sigma' _{\stackrel{\scriptstyle ab}{cd}}
  \: = \, 
\sum_{xyzt}\,
 \Gamma_{\stackrel{\scriptstyle abcd}{xyzt}}\,
\sigma _{\stackrel{\scriptstyle xy}{zt}} \ .
\label{Xmap2}
\end{equation} 
A supermap corresponds to the concept of {\sl motion},
which transforms the set  of extended pure states in
 the generalized quantum mechanics of Mielnik \cite{Mi74}.
Some properties of supermaps were independently investigated
in a very recent work by Chiribella, D'Ariano and Perinotti \cite{CDP08}.
 
Investigating linear maps in the set of extended states
we aim to accomplish two complementary tasks:
i) For any quantum operation 
  $\Psi$ construct a corresponding supermap $\Gamma=\Gamma(\Psi)$
   which preserves the set of quantum states embedded
    inside the set of extended states. 
ii) For any admissible trace-preserving supermap $\Gamma$
    which acts on the set ${\cal M}_{N}^{X}$ of extended states
    find a reduced quantum map $\Phi$,
    which acts the set ${\cal M}_{N}^{Q}$  and forms
    a quantum operation, (is completely positive and trace preserving).
 
Let us first consider the product extension of quantum operations,
\begin{equation}
\rho \to \rho'=\Psi(\rho) 
\quad \Longrightarrow \quad 
 \sigma \to \sigma' = \Gamma(\sigma)= (\Psi \otimes  {\mathbbm 1}) \sigma \ .
\label{dynam1} 
\end{equation}  
If the initial state has the product form,
$\sigma=\rho \otimes {\mathbbm 1}/N$,
then $\sigma'=\Psi(\rho)\otimes {\mathbbm 1}/N$.
The maps of the form (\ref{dynam1})
preserve thus the structure of the set of quantum states,
and in this way any quantum operation
can be realized in the extended set--up.
In this way we have arrived at

\medskip
{\bf Proposition 1}. {\sl The extended, quartic theory is a generalisation
       of the standard quantum theory. In the special case of the tensor product
   structure of initial states and supermaps, XM reduces to QM.}  
\medskip

The special case of supermaps of the product form (\ref{dynam1})
has a simple interpretation 
in view of the Jamio{\l}kowski isomorphism.
Associating by means of (\ref{jamiol2})
initial and final states, $\sigma$ and $\sigma'$, with the maps 
$\Phi$ and $\Phi'$ 
we realize that the map $\Gamma=\Psi \otimes {\mathbbm 1}$
acts in the space of extended states (identified with the set of maps)
as a {\sl composition}, $\Phi'= \Psi \cdot \Phi$.
Going back into the space of states with Jamio{\l}kowski isomorphism into the space 
of quantum states
one can define in this way a composition of states \cite{RFZ08},
$\sigma_{\Psi} \odot \sigma_{\Phi}:=(\sigma_{\Psi}^R \sigma_{\Phi}^R)^R$.

Let us now relax the assumption (\ref{dynam1}) 
on the product form and look for a more general class of
linear maps. In general we need to work with maps $\Gamma$ 
that preserve the set of extended states;
if $ \sigma \in {\cal M}_{N}^{X}$
then $\Gamma(\sigma) \in {\cal M}_{N}^{X}$.
This property parallels the positivity of quantum maps
which preserve the set ${\cal M}_{N}^{Q}$.
However, analyzing dynamics of a quantum system 
one takes into account the possible presence
of an ancilla and defines completely positive maps.
Hence we advance the following notion
of {\sl completely preserving} maps,
related to the concept of {\sl well defined} transformations
used by Barrett in \cite{Ba06}.

\smallskip
{\bf Definition}. Consider a given sequence of convex
sets $Q_k$ labeled by an integer $k$ and a map $\Gamma$
defined on $Q_N$.  The map $\Gamma$ is called 
{\sl preserving} if $\Gamma(Q_N)\subset Q_N$.
We say that the map $\Gamma$ is {\sl completely  preserving} 
if its extension $\Gamma \otimes {\mathbbm 1}_M$
acting on $Q_{NM}$ is preserving for an arbitrary $M$.
\smallskip

Taking for $Q_N$ the set  ${\cal M}_{N}^{Q}$
of quantum states we get back the standard definition of CP maps.
However, if we put for $Q_N$ the set ${\cal M}_{N}^{X}$,
we get the characterization of the maps which 
completely preserve the set of extended states,
so are admissible in the quartic theory.

It is not difficult to show that the set
of supermaps completely preserving the structure 
of ${\cal M}_{N}^{X}$ is not empty.
It contains for instance all 
maps of the product form (\ref{dynam1}),
and also all maps acting on ${\cal M}_{N^2}^{Q}$ 
which are bistochastic. Any extension of a bistochastic
map is bistochastic, and this property guarantees 
that any initial state majorizes its image,
so the  structure (\ref{restrict1}) of the set of ${\cal M}_{N}^{X}$
is preserved.

On the other hand, the problem of deciding,
whether a given map acting in the space
of extended states is preserving (completely preserving)
is in general not simple, and it is not determined by 
the (complete) positivity of the map.
For instance, a completely positive map
which sends all states of ${\cal M}_{N^2}^{Q}$ into a pure state
$|\varphi\rangle \langle \varphi | $ (where $|\varphi\rangle \in {\cal H}_{N^2}$),
is not preserving, since the pure state does not belong to  ${\cal M}_{N}^{X}$,
so it can not be completely preserving.
However, the reflection with respect to the maximally mixed state,
$\rho'=2\rho_*-\rho$,
is not positive in ${\cal M}_{4}^{Q}$,
but it preserves the smaller set ${\cal M}_{2}^{X}$
of extended states.
It is also known that allowing for a non--product extension 
followed by a global unitary dynamics
and partial trace over the auxiliary subsystem
may lead to non completely positive dynamics \cite{CTZ05}.
We close the discussion here admitting
that the problem of finding an efficient criterion
to distinguish the preserving and completely preserving maps
remains open.

In order to compare predictions of quartic and quadratic 
theory we need to find a way to associate 
with a given supermap,
$\Gamma:{\cal M}_{N}^{X} \to {\cal M}_{N}^{X}$,
a quantum map $\Psi:{\cal M}_{N}^{Q} \to {\cal M}_{N}^{Q}$.
For a moment let us restrict our attention to 
completely positive maps which act in the set 
${\cal M}_{N}^{Q}$. It can be represented 
in the standard Kraus form,
\begin{equation}
\sigma'= \Gamma(\sigma)\ =\ \sum_{i=1}^k Y_i \, \sigma  \, Y_i^{\dagger} \ .
\label{dynam2} 
\end{equation}  
The Kraus operators $Y_i$ act now in the extended Hilbert space 
${\cal H}_{N^2}={\cal H}_{A}\otimes {\cal H}_{A'}$.
In contrast to  the form (\ref{Kraus1}),
which describes all completely positive maps admissible by  standard quantum theory,
the Kraus form  (\ref{dynam2}) of a supermap 
provides only a class of measurement  processes
admissible within the extended quantum theory.

In analogy to (\ref{dynmatr3})
we may represent such a supermap $\Gamma$ by its dynamical matrix
\begin{equation}  
G \ =\ \Gamma^R \ = \  \Bigl( \sum_{i=1}^k Y_i \otimes {\bar Y_i} \Bigr)^R  \ .
\label{supermap2}
\end{equation} 
The operators $Y_i$ can be chosen to be orthogonal,
so their number $k$ will not be larger than $N^4$.
The Choi matrix $G$ of size $N^4$ is Hermitian and it acts 
on ${\cal H}_{N^4}={\cal H}_{A}\otimes {\cal H}_{A'}\otimes
{\cal H}_{B}\otimes {\cal H}_{B'}$.
Here label $A$ denotes the principal system, $A'$ its extension
which generates the state in the quartic theory,
while $B$ and $B'$ represent their counterparts
used to apply the Jamio{\l}kowski isomorphism (\ref{jamiol2}).
 
To simplify the dynamics in the space of extended states
it is enough to consider the Choi matrix obtained by 
normalizing the partial trace of the Choi matrix representing a supermap,
$D=({\rm Tr}_{A'B'}G)/N$. The resulting quantum map, 
 $\Phi=D^R$, inherits its properties
from the corresponding supermap.

\medskip
{\bf Lemma 3}. {\sl Let $\Gamma$ be a linear supermap acting on 
${\cal M}_N^X$, so it can be represented by 
a Hermitian dynamical matrix  $\Gamma^R$.
Construct a quantum map $\Psi$ acting on ${\cal M}_N^Q$
by performing the partial trace of the dynamical matrix,
$\Psi=D^R$ where $D={\rm Tr}_{A'B'}(\Gamma^R)/N$.
If ~$\Gamma$ is a completely positive (stochastic, bistochastic)
supermap, so is the quantum map $\Psi$.}
\medskip

{\bf Proof}. If a supermap $\Gamma$ is completely positive,
then due to Choi theorem the corresponding dynamical matrix is positive,
$G=\Gamma^R\ge 0$. So is its partial trace,
$ND={\rm Tr}_{A'B'}G$, which implies complete positivity of $\Psi$.
If  $\Gamma$ is trace preserving then
${\rm Tr}_{BB'}G={\mathbbm 1}_{N^2}$, hence
${\rm Tr}_{B}D={\rm Tr}_{A'}{\mathbbm 1}_{N^2}/N={\mathbbm 1}_{N}$,
which implies trace preserving condition for $\Psi$.
Analogously, if $\Gamma$ is unital then
${\rm Tr}_{AA'}G={\mathbbm 1}_{N^2}$ so
${\rm Tr}_{A}D={\rm Tr}_{B'}{\mathbbm 1}_{N^2}/N={\mathbbm 1}_{N}$
which implies unitality of $\Psi$. Thus 
stochasticity (bistochasticity)
of the supermap $\Gamma$ implies the same property of the 
associated quantum map $\Psi$. $\square$
\smallskip

If the supermap has a product form,
$\Gamma=\Psi_A \otimes \Psi_{A'}$, 
and all its Kraus operators have the tensor product
structure, the corresponding
quantum dynamics is given by reduced Kraus operators,
$X_i={\rm Tr_{A'}}Y_i$. However, in the case
of an arbitrary stochastic $\Gamma$
this relation does not hold.

In general one may also consider a wider class of supermaps
which preserve the set of extended states, but for which
the extended Choi matrix $G=\Gamma^R$ is not positive. 
However we need to require that the induced quantum dynamics
is completely positive. This implies a condition for the partial trace
\begin{equation}  
D \ = \ {\rm Tr}_{AA'}G  \ \ge \ 0  \ ,
\label{parttr2}
\end{equation} 
 which is obviously fulfilled for any positive $G$.
On the other hand relation (\ref{parttr2})
is satisfied for a large class of operators $G$ which are not positive.
This shows that the class of admissible dynamics
in the extended theory is wider than in the standard quantum theory.


\section{Classical, quantum and extended theories:\\
         a comparison}

The standard quantum theory reduces to a classical theory
if one takes into account only the diagonal parts of a 
state $\rho$ and restricts the space of operations.
Technically, one may define an 
operation of {\sl coarse--graining} with respect to a given 
Hermitian operator $H$, which is assumed to be non-degenerate.
This operation can be represented 
as a sum of projectors onto eigenstates $|h_i\rangle$ of $H$, 
\begin{equation} 
\rho \: \to \: \Phi_{CG}(\rho)  = 
\sum_{i=1}^N \, |h_i\rangle \langle h_i| \rho |h_i\rangle \langle h_i| \ . 
\label{coarse}
\end{equation}

In other words, 
this  map deletes all off-diagonal elements 
from a density matrix, if represented in the eigenbasis of $H$
and produces a probability vector
${\vec p}={\rm diag}(\rho)$. 
It consist of $N$ non-negative components,
the sum of which is not larger than unity, 
so   ${\vec p}$ lives in the 
simplex $\widetilde{\cal M}_N^C=\Delta_{N}$.
Since off-diagonal elements are called
quantum coherences,
the process induced by  coarse graining 
is called {\sl decoherence}. The effects of
decoherence  play a key role in quantum theory
and their presence explains
why effects of quantum coherence are
not easy to register.

In a similar way, for each quantum map $\Psi$
one may obtain reduced, classical dynamics, by 
taking diagonal elements of dynamical
matrix $D=\Psi^R$. The classical transition matrix,  
$T=[{\rm diag}(\Psi^R)]^R$,
inherits properties of $\Psi$, as stated in Lemma 1.
In particular, if $\Psi$ is a stochastic map,
then $T$ forms a stochastic matrix,
while if $\Psi$ is a trace non--increasing map,
then $T$ is substochastic, what means that the
sum of all elements in each its
column is not larger than unity.

Consider an arbitrary quantum state $\rho$, 
transform it by a stochastic map into $\rho'=\Psi(\rho)$
and perform coarse graining to obtain 
a classical state $p'={\rm diag}(\rho')$.
Alternatively, get the classical vector
by coarse--graining, $p={\rm diag}(\rho)$,
and transform it by reduced dynamics $T$
to arrive at $p''=Tp$.
In general, both vectors are not equal,  
\begin{equation} 
p'_m=\sum_{ab}\Psi_{\stackrel{ \scriptstyle m m}{ab}}\, \rho_{ab}
\ \ne \ 
p_m''=
\sum_{abc}\Psi_{\stackrel{ \scriptstyle m m}{ab}}\, \rho_{ac}
\delta_{ab}\delta_{ac} \ ,
\label{compCQ}
\end{equation}
which is a consequence of the known fact
that classical and quantum dynamics do differ.
Such a direct comparison between discrete classical and quantum dynamics 
may be succinctly summarized in a non--commutative diagram:
\begin{equation}
\begin{array}{llcr}
{\rm QM: \quad \quad} &
{\cal M}_N^Q \ni \rho \quad \quad &
\stackrel{\Psi}{\longrightarrow}&
   \rho'=\Psi(\rho)  \\ 
{\rm \quad} {\vphantom{\Bigg|}\downarrow }\  {\rm \quad} \Phi_{\rm CG} 
& {\rm \quad} \quad \quad {\vphantom{\Bigg|}\downarrow }   &   
{\vphantom{\Bigg|}\downarrow } &
{\vphantom{\Bigg|}\downarrow }
{\rm \quad} \Phi_{\rm CG}   \\
{\rm CM: \quad \quad} &
{\cal M}_N^C \ni p={\rm diag}(\rho) &
\stackrel{T(\Psi)}{\longrightarrow}&
  p''\ne p'={\rm diag}(\rho')  
\end{array}
\label{diagr1}
\end{equation}
Horizontal arrows represent quantum (classical) discrete dynamics, 
while vertical arrows can be interpreted as
the action of the coarse--graining operation defined in Eq. (\ref{coarse}),
which reduces quantum theory to classical.

In an analogous way one can compare dynamics
with respect to the extended and standard quantum theories.
The transition from a state of the quartic theory 
to a standard quantum mechanical state occurs by
taking the partial trace, $\rho={\rm Tr}_{A'}\sigma$.
This process can be called a {\sl hyper--decoherence},
since it corresponds to the decoherence which
induces the quantum--classical transition.  
As standard decoherence effects make the observation of the
quantum effects difficult, the hyper--decoherence
reduces the magnitude of effects unique  to the extended theory.

Let us start with an arbitrary state $\sigma$ of the quartic theory, 
transform it by an admissible linear supermap map into $\sigma'=\Gamma(\sigma)$
and perform a reduction to obtain the
quantum state $\rho'={\rm Tr}_{A'}\, (\sigma')$.
Alternatively, get the quantum state by reduction,
$\rho={\rm Tr}_{A'}\, (\sigma)$,
and transform it by reduced quantum map $\Psi(\Gamma)$,
characterized in Lemma 3, to arrive at $\rho''=\Psi(\rho)$.
In general, both quantum states are different,  
\begin{equation} 
\rho'_{mn}=\sum_{xyztb}\Gamma_{\stackrel{ \scriptstyle mbnb}{xyzt}}\, 
\sigma_{\stackrel{ \scriptstyle xy}{zt}}
\ \ne \ \
\rho''_{mn}=\sum_{xyztb}\Gamma_{\stackrel{ \scriptstyle mxny}{ztzt}}\, 
\sigma_{\stackrel{ \scriptstyle xb}{yb}} \ .
\label{compQX}
\end{equation}

In this way we have justified 

\medskip
{\bf Proposition 2}. {\sl The extended quartic theory forms a nontrivial
generalisation of the standard quantum theory. In particular,
there exist experimental schemes (consisting of an initial state
and the measurement operators) for which 
both theories give different predictions concerning probabilities 
recorded.}  
\medskip

The comparison between dynamics in quartic and quadratic theories
is  visualized in  a non--commutative diagram analogous to 
(\ref{diagr1}),
\begin{equation}
\begin{array}{llcr}
{\rm XM: \quad \quad} &
{\cal M}_N^X \ni \sigma \quad \quad &
\stackrel{\Gamma}{\longrightarrow}&
   \sigma'=\Gamma(\sigma)  \\ 
 {\vphantom{\Bigg|}\downarrow }\  {\rm \quad reduction}  
& {\rm \quad} \quad \quad {\vphantom{\Bigg|}\downarrow }   &   
{\vphantom{\Bigg|}\downarrow } &
{\rm partial ~ trace~} {\vphantom{\Bigg|}\downarrow }
{\rm \quad}  \\
{\rm QM: \quad \quad} &
{\cal M}_N^Q \ni \rho={\rm Tr}_{A'}(\sigma) &
\stackrel{\Psi(\Gamma)}{\longrightarrow}&
  \rho''\ne \rho'={\rm Tr}_{A'}(\sigma')  
\end{array}
\label{diagr2}
\end{equation}
The vertical arrows denote here the operation of 
partial trace over auxiliary subsystem and
reduction of  extended theory to quantum theory, while  
horizontal arrows represent dynamics
in the space of extended (quantum) states.

Thus the classical theory describing dynamics inside 
the $N$ dimensional simplex $\Delta_{N}$ of 
subnormalized probability vectors 
remains a special case of the quantum theory,
in which dynamics takes place in $N^2$ dimensional 
set $\widetilde{\cal M}_N^Q$ of subnormalized quantum states.
In a very similar manner, the standard quantum theory
may be considered as a special case
of the extended theory, obtained by projecting down 
the $N^4$ dimensional set  $\widetilde{\cal M}_N^X$
of extended states into $\widetilde{\cal M}_N^Q$.
A comparison between all three approaches is summarized
in Table \ref{tab1}. The symbol ${}^R$ denotes here
the transformation of reshuffling of a matrix defined in  Eq. (\ref{dynmatr3}).

\begin{table}
\caption{Comparison between classical, quantum and extended theory.}
  \smallskip
{\renewcommand{\arraystretch}{1.57}
\begin{tabular}
[c]{l c c  c}\hline \hline
Theory  
& \parbox{2.8cm}{\centering Classical \\ (linear)} 
& \parbox{2.8cm}{\centering Quantum \\ (quadratic)} 
& \parbox{2.8cm}{\centering Extended  \\ (quartic)} 
\\ \hline
pure states
& corners of $\Delta_{N-1}$ &
$|\psi\rangle \in {\mathbb C}{\bf P}^{N-1}$ &
$
U(|\psi\rangle \langle \psi| \otimes 
\frac{{\mathbbm 1}}{N})U^{\dagger} $ \\
mixed states
&
 ${\vec p} \in \widetilde{\cal M}_N^C=\Delta_{N}$ &
$\rho \in \widetilde{{\cal M}}_N^Q$  &
$\sigma \in \widetilde{\cal M}_N^X$ \\
dimensionality & $N$ & $N^2$ & $N^4$ \\
dynamics & ${\vec p}\, '=T{\vec p}$ &
$\rho'=\Psi(\rho)$ &
$\sigma'= \Gamma(\sigma)$ \\
 \hline
\end{tabular}
\begin{tabular}
[c]{l l l}
 \parbox{3.4cm}{ reduction \quad from }  &
\parbox{3.6cm}{quantum to classical}  &
\parbox{4.0cm}{extended to quantum}  \\
\hline
of the state &
\quad \quad $p={\rm diag}(\rho)$ &
\quad $\rho={\rm Tr}_{A'}\, \sigma $   \\
of the map  &
\quad  $T=[{\rm diag}(\Psi^R)]^R$ &
 $\Psi=\bigl[{\rm Tr}_{A'B'} \Gamma^R\bigr]^R/N$  \\
\hline \hline
\end{tabular}
}
\label{tab1}
\end{table}  

To reveal similarities between both decoherence processes let us 
formulate two analogous statements.

\medskip
{\bf Proposition 3: Decoherence.} 
 {\sl  A classical state $\vec p'$ obtained by
a decoherence of an 
arbitrary quantum state corresponding to the classical state
$\vec p$ satisfies the majorization relation
\begin{equation} 
{\vec p}' \ := \ {\rm diag}(UpU^{\dagger}) \ \prec \  {\vec p}
\label{decoh1} 
\end{equation}
where $U$ is a unitary matrix of size $N$  
and  $p$ stands for a diagonal matrix with vector $\vec p$ on the diagonal.}
\smallskip

The proof consists in an application of the Schur lemma, 
which states that the diagonal of a positive Hermitian matrix
is majorized by its spectrum. This statement follows also 
from the  Horn--Littlewood--Polya lemma,
which says that ${\vec x} \prec {\vec y}$
if there exists a bistochastic matrix $B$
such that $x=By$ -- see e.g.  \cite{BZ06}.
\medskip

{\bf Proposition 4: Hyper--decoherence.}  
{\sl  A quantum  state $\rho'$ obtained by
a hyper--decoherence from an arbitrary extension 
of a quantum state $\rho$ 
satisfies the majorization relation
\begin{equation} 
\rho ' \ := \ {\rm Tr}_{A'} 
[U (\rho \otimes {\mathbbm 1}/N) U^{\dagger}] \ 
\prec \  \rho 
\label{decoh2} 
\end{equation}
where $U$ is a unitary matrix of size $N^2$.}
\smallskip

To prove this statement it is enough to observe that
the map $\rho'=\Phi(\rho)$ is bistochastic, 
so according to the quantum analogue of the
Horn--Littlewood--Polya lemma (see e.g.  \cite{BZ06}),
the majorization relation (\ref{decoh2}) holds.
Alternatively, for small system sizes one may 
use  results of the quantum marginal problem: 
inequalities of Bravyi \cite{Br04} for $N=2$
and inequalities of Klyachko \cite{Kl04} for $N=3$
concerning constraints  between the spectra
of a composite system and its partial traces
imply relation (\ref{decoh2}).

Note that the unitary matrix $U$ is arbitrary,
it may in particular represent the swap operation,
which exchanges both subsystems.
Thus the extended state
should not be treated as a merely composition 
of a 'physical particle' with an 'auxiliary ghost':
They are intirinsicly  intertwined
into a single entity representing an extended state $\sigma$,
which may be reduced to the standard quantum state $\rho'$
due to the process of hyper--decoherence.

\medskip

\section{Higher order theories}

Iterating the extension procedure one can construct 
higher-order theories, in which the number of degrees
of freedom scales with dimensionality as $K=N^r$
for any even $r$.  Let us rename the set  
$ {{\cal M}}_{N}^{Q}$
of quantum states into   $ {{\cal M}}_{N}^{(0)}$,
and the set $ {{\cal M}}_{N}^{X}$
of extended states  into $ {{\cal M}}_{N}^{(1)}$.
Then we may define the set of states in a $m$-th order
generalized theory  as in (\ref{restrict1}),  
\begin{equation} 
{{\cal M}}_{N}^{(m)} := \{\sigma\in {\cal M}_{N^{1+m}}^{(0)}:  
 \sigma \prec \sigma_{0}=
|0 \rangle \langle 0 | \otimes \frac{1}{N^m}{\mathbbm 1}_{N^m} \} \ ,
\label{restrictk} 
\end{equation}
where $|0\rangle \in {\cal H}_N$.
The parameter $m=1,2,3\dots$ represents the number of 
additional ancilliary states: It is equal to zero for the
standard quantum theory, 
$m=1$ for the extended quartic theory discussed
in this work, and $m=2$ for the next,
extended quantum  theory for which $K=N^6$.
In general, the number $K$ of degrees of freedom 
behaves as $N^{2m+2}$,
so the exponent $r$ is equal to $2m+2$.
The standard quantum state is obtained by
the partial trace over an 
auxiliary system of size $N^m$.

The spectra of the states of the
higher order theories also form  a
permutohedron
${\rm Perm}(\{N^{-m}\}_{N^m} , \{0\}_{N^m(N-1)})$,
 defined by the vector of length $N^{m+1}$
containing  $N^m$ non--zero equal components.
From a geometric point of view increasing the number
$m$ of ancillas corresponds to increasing
the dimensionality of the Hilbert space and continuing the
procedure of truncation of the set of positive operators,
which represent quantum states. The larger $m$,
the more faces and corners of the permutohedron
which becomes closer to the ball
in  $N^{2m+2}$ dimensions.

The set of pure states of the higher order theory forms the
flag manifold
${\cal P}_N^{(m)}=U(N^{m+1}) / [(U(N^{m+1}-N)\times U(N)]$
of  $2(N^{m+2}-N^2)$ dimensions. 
The entropy of any state of such a theory belongs
to the interval $S(\sigma)\in [m\ln N, (m+1)\ln N]$,
so its degree of mixing can be characterized by 
the gauged entropy,  $S_m:=S-m\ln N$,
which is equal to zero for extended pure states.

As analyzed in section section \ref{sec:POVM}
the set of elements of extended POVM
can be defined by the cone dual to the set of extended states,
${\cal E}_N^X = (\widetilde{{\cal M}}_N^{X})^{*}$.
In a similar  way,  positivity condition  analogous to (\ref{trace2}) 
implies that  for an extended theory of order $m$
elements of  extended POVMs belong to the set 
\begin{equation} 
{\cal E}_{N}^{(m)}  \ = \   \{
E_i=E_i^{\dagger}: \ 
 {\rm eig} (E_i)  \in 
 \bigl( {\cal M}_{N}^{(m)}\bigr)^* 
 {\rm \quad and \quad} E_i \le {\mathbbm 1}_{N^{m+1}}  \}  \ .
\label{POVMb} 
\end{equation}

The larger $m$, the larger space one works with and 
the smaller (more truncated) set of extended states.
Accordingly, the larger is the corresponding dual 
set of elements of extended POVMs.

Discrete dynamics in higher order theories can be defined
as in the quartic theory. To get a reduction of 
a hypermap acting in the space of extended states
it is sufficient to perform a suitable partial trace
on an extended dynamical matrix
and require that the standard Choi matrix $D$
obtained in this way is positive.
In this manner one can define an entire hierarchy of hyper-maps
which act on the spaces of states of different dimensionalities. 

Higher order theories have an appealing feature
if the exponent is a power of two, $r=2^k$.
Then one may consider a higher order Jamio{\l}kowski isomorphism:
A hypermap acting in the $N^r$
dimensional space of extended states can be considered
as an extended state of the higher order theory
characterized by the exponent $2r$.
For instance,  a supermap $\Gamma$ of the quartic theory, $k=2$,
may be interpreted as an extended state in an octonic theory
for which $k=3$.
The state $\sigma_8=\rho\otimes {\mathbbm 1}_{N^3}/N^3 \in {\cal M}_N^{(3)}$
describes a supermap $\Gamma_{\rho}$,
which sends all quantum maps 
into $\Phi_{\rho}$  defined in (\ref{extend2}).

We have thus shown that the standard quantum mechanics
can be embedded in an infinite onion--like structure
of higher order theories.
Working with a theory characterized by an exponent $r$
one needs $N^r$ parameters to describe the quantum state
and to predict the future.
Although it is yet uncertain, whether such higher--order theories 
may have a direct physical relevance, investigating 
how one theory is embedded into another one
may contribute to a better understanding
of the structure of the standard quantum theory.

\section{Concluding remarks}
In his 1974 paper on generalized quantum mechanics \cite{Mi74}
Mielnik wrote: 
{\sl The incompleteness  of the present day science at this point is,
perhaps, one more reason why the scheme of quantum mechanics 
should not be prematurely closed.}
Although more than thirty years have gone,
we believe that this statement is still valid and it
provides motivation for the present work.

Making use of a geometric approach to quantum mechanics
we considered different constructions of the 
arena for an extended quantum theory.
In particular for the case of mono-partite systems
 we have formulated a generalisation of the standard quantum theory 
for which the number of degrees
of freedom scales as the fourth power of the 
number of distinguishable states.
As the standard quantum theory is somehow coded in the shape of the $N^2-1$
dimensional convex set ${\cal M}_{N}^Q$ of all normalized states acting on $N$
dimensional Hilbert space, the same is true for the extended theory
which deals with $N^4-1$ dimensional set  convex ${\cal M}_{N}^X$
of extended states.
It is tempting to believe that the present paper allows one to
formulate a long--term research project on 
plausible generalisations of the quantum theory and its
interdisciplinary implications.

The extended theory proposed in this work is based on the following steps:

\noindent
i) {\sl extended states} belong to the set ${\cal M}_N^X$ which  
   forms a part of the set ${\cal M}_{N^2}^Q$ of the states of bipartite system
   of the standard quantum theory truncated in such a way to support $N$ 
    distinguishable states.

\noindent
 ii) {\sl extended measurements} are formed by elements $E_i$ of the set 
   ${\cal E}_N^X$ dual to ${\cal M}_N^X$. 

\noindent
 iii) {\sl extended dynamics} is described by linear maps preserving the set 
      of extended states for which the reduced dynamical matrix is positive (\ref{parttr2}).

\medskip

The states of the extended quartic theory 
admit a dynamical interpretation
in view of the Jamio{\l}kowski isomorphism  \cite{Ja72}. 
Any extention $\sigma \in {\cal M}_{N}^{X}$
of the quantum state $\rho \in {\cal M}_{N}^{Q}$
represents a completely positive map 
  $\Phi:\ {\cal M}_{N}^{Q} \to {\cal M}_{N}^{Q}$,
which sends the maximally mixed state into 
the state in question, $\Phi({\mathbbm 1}/N)=\rho$.
Geometric approach explored in this work
allows us to construct an infinite hierarchy of higher order
theories and to extend the Jamio{\l}kowski isomorphism:
any state of the theory of order $k+1$
can be interpreted as a map acting on the
set of  states defined in the theory of order $k$.  

The extended quartic theory includes the quadratic  
quantum theory as a special case: 
If the initial state and the measurement
operators  have the tensor product form
both theories yield exactly the same results.
More formally one can assume that the entire system
in the framework of the extended  theory can be 
approximated by the Hamiltonian
\begin{equation} 
H \ \approx \ H_{\rm phys} + H_{\rm  gh}  + \lambda H_{\rm int} \ ,
\label{Ham1} 
\end{equation}
in which  $H_{\rm phys}$ and  $H_{\rm  gh}$
represent the 'physical particle' and the ancilliary subsystem, respectively,
while $H_{\rm int}$ describes the interaction between them.
The effective coupling constant $\lambda$ can be then defined
by the ratio of the expectation values,
$\langle H_{\rm int}\rangle /  \langle  H_{\rm phys} \rangle$.
The correspondence principle is obtained taking the limit $\lambda \to 0$,
since in this case the presence of the ancilliary system will not influence 
the physical system. 
On the other hand, the quartic theory is more general than the standard quantum theory 
and for a non-zero coupling constant $\lambda$ it may give different predictions 
for the probabilities of outcomes of certain measurements.

A reader will note that the structure of composite systems 
 is not touched in this scheme, which right now concerns simple
systems only. Generalisation of the extended theory
for composite systems turns out  not to be simple.
For instance, accepting the tensor product structure used in quantum theory,
one faces the problem that the same mathematical notion of partial trace
has to be used to trace out the fictitious 'ghost particle' of an extension
or a part of the physical subsystem. In particular,
an extended state $\rho_{ABA'B'}$ of two qubits, 
reduced by the partial trace with respect to the ancillas $A'B'$
can form an arbitrary two qubit state 
$\rho_{AB}={\rm tr}_{A'B'} (\rho_{ABA'B'}) \in {\cal M}_{16}^Q$,
while the same state reduced by one subsystem 
gives an extended state $\rho_{AA'}={\rm tr}_{BB'}(\rho_{ABA'B'})$
which is required to belong to a smaller set ${\cal M}_2^X \subset {\cal M}_4^Q$.
This may suggest that the extended state of two qubits 
${\cal M}_{2,2}^X \in {\cal M}_{16}^Q$ cannot be invariant with respect to the
unitary  group $U(16)$,
so the standard tensor product rule, used in quantum mechanics
to describe composed systems, cannot be 
directly adopted for the extended quartic theory.

The question whether the extended theory proposed here
may be generalized for composite systems and might
describe the physical reality remains open.
Let us mention here that composite systems 
also pose a difficulty in nonlinear generalizations 
of quantum theory.  
On the other hand, it is also thinkable to treat 
the entire Universe as the only physical system,  
and to agree that splitting it into various subsystems
is performed for practical purposes as a kind of an inevitable approximation.

Even concerning single--particle systems, it remains as a challenge to 
propose an experimental scheme in which predictions of the 
extended (quartic) theory constructed in this work
would differ significantly  from results following 
from the standard (quadratic) quantum theory.
The key issue is to find a way
to prepare a non-standard initial state $\sigma$ 
which is not in a product state but
reveals an entanglement with the ancilla.
A  suitably chosen detection scheme could then reveal 
usefulness of the generalized quartic mechanics, 
and detect the presence of the hypothetical 
ancilliary state associated with this theory.

If this step turns out not to be realistic,
by analyzing experimental data one can try
to get an upper bound for the time of the hyper-decoherence:
it is thinkable that in standard experimental conditions
such hyper-decoherence effects occur so fast
that fine effects due to correlations with 
the auxiliary system cannot be observed.
However, this could be the case for systems in extremal conditions
like these characterized by very high energies or ultrastrong fields
and important for the investigations of the early history of the universe.

Construction of an extended, quartic theory
might have consequences for the 
axiomatic approach to quantum mechanics. 
Since the proposed set up does not include composite systems, 
the present work does not imply that 
the last, simplicity axiom proposed by Hardy \cite{Ha01} 
is inevitable to derive the standard quantum theory in an axiomatic way. 
It is also plausible that this axiom is not necessary, if the
quantum theory is the only theory which satisfies all other axioms
including the one on description of composite systems.

Our work provides a starting point in attempt
to construct a fruitful generalization of the standard 
quantum theory and leaves many questions open.
One could look for a suitable algebraic viewpoint to 
study properties of the quartic (higher order)  quantum theory
or to investigate possibilities to construct an extended 
quantum theory of fields. For instance, the  'ghost--like' ancillary subsystem, 
used to interpret the extended quartic theory, 
could correspond in the path integration approach to
extending the path of integration by two additional 'ghost--like' points,
which vanish in the standard theory.
Alternatively, one could examine, if such an extended quantum theory
is related to the 'thermo field  dynamics' of Umezawa et al.
in which a 'thermal vaccum state' is introduced \cite{AU87}.

Furthermore, it would be interesting to study a
possible link between the higher order theory and 
the generalized quantum mechanics of Sorkin to verify, whether
the higher order interference terms, 
present in the generalized measure theory \cite{So94,MOS05},
are related to the extended quantum theory.
Although the quartic quantum theory is linear,
following the approach  of Mielnik \cite{Mi74} 
one could also analyze possible relations
to certain non-linear generalizations of quantum theory 
-- see e.g. \cite{We89,CD02,Go08}.

The theory of information processing can be studied
not only within the classical or the quantum set-up,
but also in more general probabilistic theories
\cite{BLMPPR05,SPG06,Ba06,Barn06,BBLW06,BBLW07,BDLT08}.
Hence one could also attempt to analyze implications
of the extended quartic theory for the 
quantum information processing.
During the last two decades it has been investigated
to what extend the transition from classical bits
to quantum qubits gives additional
possibilities for information processing.
In a similar way, one could study consequences of a
further enlargement of the scene for
an information processing screenplay.

\medskip

I am indebted to L. Hardy for numerous 
inspiring discussions and hospitality at
Perimeter Institute for Theoretical Physics
where this work was initiated.
It is also a pleasure to thank H.~Barnum, I.~Bengtsson,
I.~Bia{\l}ynicki--Birula, P.~Busch, B.~Englert,  M.~Fannes, L.~Freidel, C.~Fuchs,  
D.~Gottesman, M.~Horodecki, P.~Horodecki, J.~Kijowski, 
A.~Kossakowski, M. Ku{\'s},  M.~Leifer, D.~Markham,  J.~Miszczak, 
G.~Sarbicki, R.~Sorkin, St.~Szarek, D.~Terno and A~Vourdas for helpful remarks and stimulating discussions.
I acknowledge financial support by the special
grant number DFG-SFB/38/2007 of Polish Ministry of Science 
and by the European research program COCOS.

\appendix
\section{Dual cones and dual sets}

In this appendix we provide necessary concepts
of convex analysis and prove a lemma on dual sets.

Consider any set $C$ in ${\mathbbm R}^n$.
Its {\sl dual cone} is defined as \cite{La07}
\begin{equation}  
C^* \ : = \  \bigl\{ y  \in {\mathbbm R}^n\ : \ \  x\cdot y \ge 0,   \forall x \in C \bigr\} \ .
\label{dual1}
\end{equation} 
The set $C^{**}=(C^{*})^{*}$ is the closure of the smallest cone 
containing $C$. A cone $C$ is said to be {\sl self--dual} if $C=C^*$.
The non-negative orthant of ${\mathbbm R}^n$ and the space
of all positive semi--definite matrices are self--dual.
 
Normalization condition $\sum_{i=1}^n x_i=1$
defines a hyperplane $H_1$ in  ${\mathbbm R}^n$.
Let $V$ be an arbitrary set 
obtained as a cross-section of a convex cone $C$,
with the hyperplane, $V=C\cap H_1$.
Then its {\sl dual set} 
is given by the cross--section of the hyperplane with the dual cone,
$V^*= C^* \cap H_1$ -- see Fig. \ref{figa1}.
Note that the probability simplex is self-dual, $\Delta_N=\Delta_N^*$.
Consider, for instance, a symmetric subset of the $N=2$ simplex,
$V_a=[a,1-a]$ with $a\in [0,1/2]$. Then the dual set reads
$V_a^*=[b,1-b]$ where $b=a/(2a-1)$. Hence 
if $V$ reduces to a single point for $a=1/2$,
its dual $V^*$ covers entire line.
Two more examples of pairs of dual sets,
 which live in the plane $H_1$ defined by the condition  $x_1+x_2+x_3=1$
are shown in  Fig. \ref{figa2}.

\begin{figure} [htbp]
      \begin{center} \
  \includegraphics[width=10.0cm,angle=0]{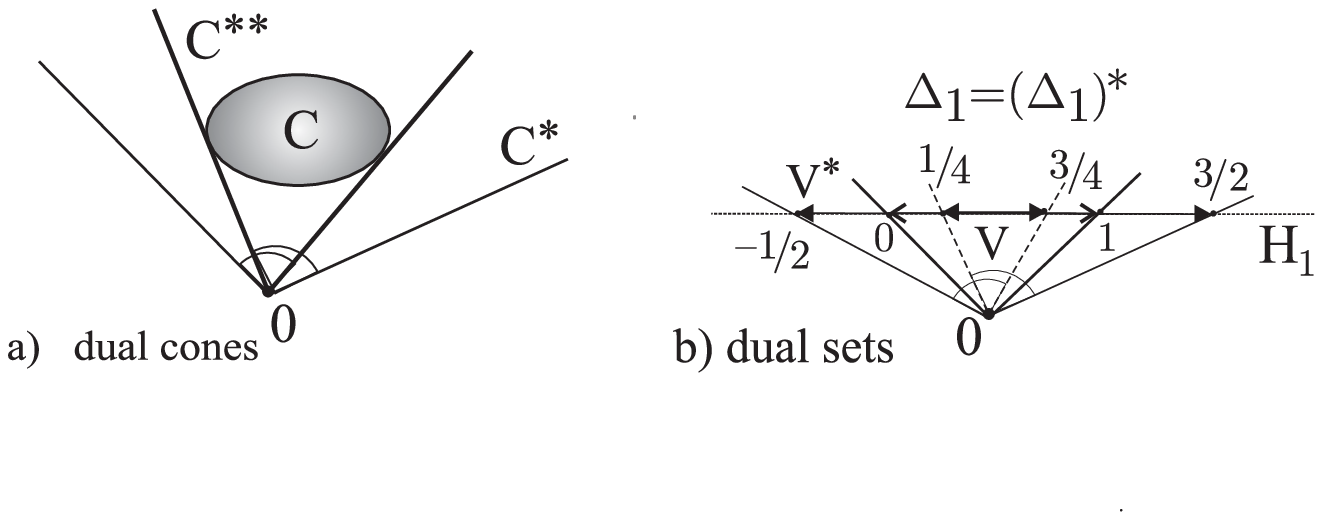}
\caption{Dual cones and sets:  a) a set $C$, its dual cone $C^*$,
and the cone $C^{**}\supset C$;
 b) a set $V_a\subset \Delta_1$ and its dual set $V_a^*\subset H_1$ plotted here for $a=1/4$.}
\label{figa1}
\end{center}
    \end{figure}

The same concept of dual sets can be applied for the set
of Hermitian operators. 
Instead of the standard scalar product one uses in this case
the Hilbert--Schmidt product, 
$\langle A|B \rangle={\rm tr} A^{\dagger}B$,
while the hyperplane is introduced by the trace normalization,
$A\in H_1 \Leftrightarrow A=A^{\dagger}, {\rm tr} A=1$.
Hence for any set $V$ in $H_1$ its dual reads
\begin{equation}  
V^* \ : = \  \bigl\{ A  \in H_1\ : \ \ {\rm tr} AB  \ge 0, \ \   \forall B  \in V \bigr\} \ .
\label{dual2}
\end{equation} 

Taking a brief look at eq (\ref{trace2})
we see that the cone containing the set  
${\cal M}_N$ of extended states and the cone 
${\cal E}_N$ of admissible elements of a POVM have to be dual.
In the standard quantum theory (quadratic), 
these cones are selfdual,  $({\cal M}_N^Q)^* ={\cal E}_N^Q$. 
However, this is not the case for extended (quartic) theory:
Since  ${\cal M}_N^X$ is obtained by {\sl truncating}
the set of positive operators as discussed in section 4,
its dual cone  $({\cal X}_N^X)^*$ 
is {\sl extended} to contain  also some non-positive operators.

The structure of the set dual to some set of Hermitian matrices gets simpler if we consider a class
of sets, which are invariant with respect to all unitary operations,
$A\in V_Q \Rightarrow UAU^{\dagger} \in V_Q$.
In such a case one can reduce the problem of finding the dual set $X^*$
of operators to a simpler problem of finding the set 
dual to the set of all admissible spectra.

\medskip

\begin{figure} [htbp]
      \begin{center} \
  \includegraphics[width=10.0cm,angle=0]{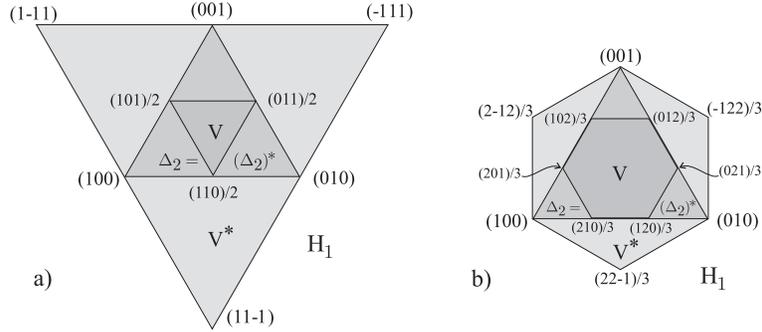}
\caption{Selfdual simplex $\Delta_2=\Delta_2^*$ of $N=3$ classical states
  living in the plane $H_1$ and its subsets:
   a) triangle  $V={\rm Perm}(1/2,1/2,0)$  and its dual triangle $V^*={\rm Perm}(1,1,-1)$;
    b) hexagon $V={\rm Perm}(2/3,1/3,0)$  and its dual hexagon $V^*=
      {\rm conv \ hull}\bigl({\rm Perm}(2/3,2/3,-1/3) \cup 
                                    {\rm Perm}(1,0,0) \bigr)$.}
\label{figa2}
\end{center}
    \end{figure}
\medskip

{\bf Lemma 4.}
{\sl Let $V$ be a convex subset of the simplex $\Delta_{N-1}$ of classical probability 
vectors of length $N$, which is invariant with respect to all $N!$ permutations
of the vector components. Let ${ V}_Q$ denote the set of all Hermitian matrices 
$UvU^{\dagger}$, where $U$ is unitary and $v$ is a diagonal matrix of size $N$
such that ${\rm diag}(v)\in { V}$.
Then the dual set $({V}_Q)^{*}$ 
contains all Hermitian operators with spectra belonging to $V^*$, so it can be called
$({V}^{*})_Q$.}  

\bigskip

Observe that this lemma can also be visualized by  Fig. \ref{figa2}:
If two sets of classical normalized vectors  $V$ and $V^*$ from $H_1$
are dual, so are the sets $V_Q$ and $(V^*)_Q$ 
of Hermitian operators with spectra belonging to $V$ and $V^*$, respectively.
Before presenting its proof let us quote a related  lemma
proved in \cite{MMZ07}.   

\medskip

{\bf Lemma 5.}
{\sl
Consider a state $\rho\ge 0$ with spectrum $\vec p$
and a Hermitian operator $\sigma=\sigma^{\dagger}$
not necessarily positive, with spectrum $\vec q$. Then
their trace is bounded as}
 \begin{equation}
p^{\uparrow} \cdot q^{\downarrow}
\  \le \    {\rm Tr}\rho \; \sigma  \   \le \
p^{\uparrow} \cdot q^{\uparrow}  \ ,
\label{tracest2}
\end{equation}
{\sl where $p^{\uparrow}$ ($p^{\downarrow}$)
denotes vectors in an increasing (decreasing) order.}

\medskip

{\bf Proof of Lemma 4.}
Lemma 5 applied to a positive operator $A \in {V}_Q$ 
and a Hermitian operator $B=B^{\dagger}$ with spectrum $\vec q$  gives 
a lower bound for the trace ${\rm tr} AB$ in terms of their ordered spectra.
Since ${\vec p}= {\rm eig}(A) \in {V}$ 
so that if $\vec q$ belongs to the dual set ${V}^*$
their scalar product ${\vec p} \cdot {\vec q}$ is non-negative.
This relation holds for any order of the components of both vectors,
since $V$  is invariant with respect to permutations.
This fact implies in turn that  ${\rm tr} AB\ge 0$ for
any hermitian operator $B$ with spectrum ${\vec q}$,
which is equivalent to the statement $B \in  ({V}^*)_Q$.
In this way we have shown that the set dual to  
 ${V}_Q$ is equal to $({V}^{*})_Q$. $\square$

\medskip

In particular  Lemma 4 implies 
relation (\ref{povme2}) which shows
that the set ${\cal E}_N^X$ of elements of XPOVM contains 
Hermitian operators with spectra from the cone
dual to the set of spectra of all extended states ${\cal M}_N^X$.
Furthermore, for any extended theory of oder $m$
it follows that  relation (\ref{POVMb}) holds and 
the cones of extended states and elements of extended POVMs are dual.

\end{document}